# Alloying V in $MnBi_2Te_4$ for Robust Ferromagnetic Coupling and Quantum Anomalous Hall Effect

Y. S. Hou[1] and R. Q. Wu[1]

[1] Department of Physics and Astronomy, University of California, Irvine, CA 92697-4575, USA

**Abstract**

The intrinsic antiferromagnetic (AFM) interlayer coupling in two-dimensional magnetic topological insulator $MnBi_2Te_4$ places a restriction on realizing stable quantum anomalous Hall effect (QAHE) [Y. Deng *et al*., Science 367, 895 (2020)]. Through density functional theory calculations, we demonstrate the possibility of tuning the AFM coupling to the ferromagnetic coupling in $MnBi_2Te_4$ films by alloying about 50% V with Mn. As a result, QAHE can be achieved without alternation with the even or odd septuple layers. This provides a practical strategy to get robust QAHE in ultrathin $MnBi_2Te_4$ films, rendering them attractive for technological innovations.

Email: wur@uci.edu

The combination of nontrivial band topology and long range magnetic ordering in magnetic topological insulators (TIs) may lead to various exotic quantum states, such as quantum anomalous Hall effect (QAHE), topological superconductivity and Majorana state [1,2]. This opens vast opportunities for advancing the next-generation quantum technologies [3]. The recently discovered van der Waals (vdW) antiferromagnetic (AFM) TI, MnBi$_2$Te$_4$ (MBT), possesses topological and magnetic orders simultaneously [4-8] and is viewed as a promising candidate for the realization of dissipationless QAHE at a reasonably high temperature [6]. The long range ferromagnetic (FM) ordering of spin moments of Mn within the septuple layers (SLs) and the subsequently induced large band gap of several-ten milli-electron volts at its topological surface states (TSSs) [4,5,9-11] permit having QAHE at much higher temperatures than in 3$d$ transition metal doped TIs [12-14]. However, due to the AFM interlayer coupling between the FM SLs, QAHE is realized in MBT thin films only with odd numbers of SLs [6,15-17]. Such thickness-dependent occurrence of QAHE in MBT thin films poses an outstanding challenge on fabrication procedures for practical applications. Furthermore, it was experimentally shown that the anomalous Hall conductivity of an odd-SL MBT thin film quantizes exactly to be $\sigma_{xy} = e^2/h$ (here $e$ is the charge of an electron and $h$ is the Planck constant) only when its magnetization is fully aligned by a large external magnetic field [6]. Therefore, it is crucial to establish stable FM interlayer coupling in MBT thin films without damaging the nontrivial band topology and intralayer FM ordering for practical utilizations of the QHAE.

In this Letter, we demonstrate through systematic density functional theory (DFT) calculations that alloying V in MBT, thus forming Mn$_{1-x}$V$_x$Bi$_2$Te$_4$, can tune the AFM interlayer coupling to the FM one in a wide range of $x$. In addition, Mn$_{1-x}$V$_x$Bi$_2$Te$_4$ also has a higher Curie temperature than the odd-SL MBT thin films. Consequently, thickness-independent QAHE can show up in Mn$_{1-x}$V$_x$Bi$_2$Te$_4$ thin films with a further elevated temperature. Our work establishes the V-alloyed MBT magmatic TI as a promising platform to make use of the dissipationless QAHE in next-generation topological electronic devices.

Our DFT calculations are carried out by using the Vienna *Ab initio* Simulation Package with the projector-augmented wave pseudopotentials [18,19]. The PBEsol functional [20] is adopted here because it gives rather good Wyckoff positions and lattice constants for the bulk MBT [21] (Table S1 in supplementary materials (SM) [22]). We choose an energy cutoff of 350 eV and relax all structures with a force convergence criterion of 0.01 eV/Å. To consider the strong correlation effect among 3$d$ electrons, we utilize $U_{Mn}$=6 and $J^H_{Mn}=1$ eV for Mn [23,24] and $U_V$=4 and $J^H_V=1$ eV for V [17]. Nevertheless, our main results are independent of $U_{Mn}$ and $U_V$ (Part II in SM). The nonlocal vdW functional of optB86b-vdW [25,26] is used to describe the interlayer vdW interactions. The Chern number, $C_N$, and one-dimensional (1D) chiral edge states are obtained by using Wannier90 [27] and WannierTools [28].

*Table I. Here listed are intralayer and interlayer interaction parameters (in the unit of meV). Negative (positive) parameters mean FM (AFM) interactions.*

| Intralayer | $J_1$ | $J_2$ | $J_3$ | Interlayer | $J_{1,z}$ | $J_{2,z}$ | $J_{3,z}$ |
|---|---|---|---|---|---|---|---|
| MBT SL | -1.72 | 0.07 | -0.05 | 2-SL MBT | 0.05 | 0.06 | 0.00 |
| VBT SL | -3.95 | 0.22 | 0.20 | MBT SL/VBT SL | -0.03 | -0.06 | -0.02 |
| one V in MBT SL | -2.93 | 0.23 | 0.15 | | | | |

It is known that the tetradymite-type MBT bulk is an AFM TI (Fig. 1a) [4,6,29]. Within its stacking unit, the Te-Bi-Te-Mn-Te-Bi-Te SL, the Mn$^{2+}$ ion is surrounded by a distorted octahedron of six Te$^{2-}$ ions and has an electronic configuration of $t_{2g}^{\uparrow,3}e_g^{\uparrow,2}$. To describe the magnetic properties of MBT, we employ the following Heisenberg Hamiltonian [4,30]:

$$H = \sum_{ij\in\text{interlayer}} J_{ij} S_i \cdot S_j + \sum_{ij\in\text{intralyer}} J_{ij,z} S_i \cdot S_j + \sum_i A_i \left(S_i^z\right)^2 \qquad (1).$$

In Eq. (1), $J_{ij}$, $J_{ij,z}$ and $A_i$ are intralayer and interlayer interaction parameters (Fig. 1b)

and single ion anisotropy energy, respectively. Our calculations show that MBT SL has a FM nearest-neighbor (NN) intralayer interaction ($J_1$), which dominates over the second-neighbor ($J_2$) and third-neighbor ($J_3$) ones (Table I), consistent with previous studies [17,30]. Besides, the interlay exchange interactions between Mn ions in adjacent SLs are AFM (Table I), also in accordance with experiments [4,6,29]. As a result of the interlayer AFM ordering and the nontrivial band topology, MBT bulk is an AFM TI [7,17] and QAHE is achieved only in its odd-SL thin films [6,11,16].

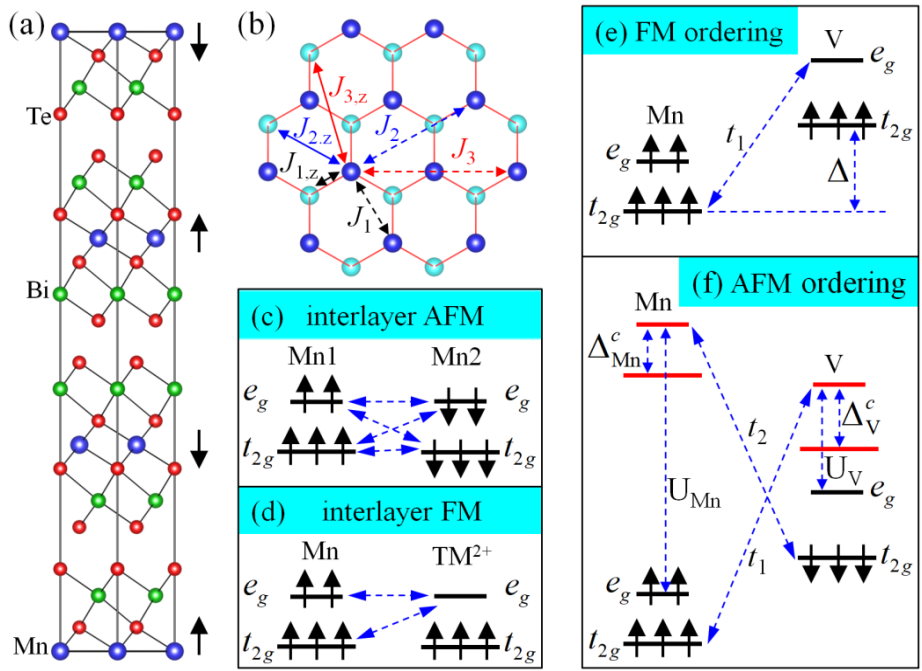

*Fig. 1. (a) Crystal structure of MBT bulk. The interlayer AFM ordering is sketched by the blacked arrows. (b) Top view of the Mn sub-lattice in 2-SL MBT and the intralayer (dashed double arrows) and interlayer (solid double arrows) interaction paths. (c) Allowed hoppings in the interlayer AFM ordering in MBT. (d) Allowed hoppings between $Mn^{2+}$ and $TM^{2+}$ ions with an interlayer FM ordering. (e) and (f) show the allowed hoppings between Mn and V in the FM and AFM orderings, respectively. Energy levels are roughly aligned based on the density of states of $Mn_{0.5}V_{0.5}Bi_2Te_4$ SL (Fig. S5 in SM).*

As the key step to remove the alternation of the QAHE in MBT thin films, we explore the mechanism of the AFM interlayer coupling. Although the interactions between interlayer Mn pairs are beyond superexchange or super-superexchange theories [31,32] due to their long separations (13.9 Å at least) across six layers of Bi and Se, they can be understood as follows. In the AFM ordering (Fig. 1c), hoppings of $t_{2g}$-$t_{2g}$, $t_{2g}$-$e_g$ and $e_g$-$e_g$ via $p$ orbitals of the intermediate Te and Bi atoms are allowed in principle, hence giving rise to an energy gain. However, these hopping paths are blocked in the FM ordering and hence the interlayer AFM ordering prevails. The situation may change if there are empty $e_g$ orbitals, as hoppings are allowed between the occupied $t_{2g}$ and $e_g$ orbitals and the empty $e_g$ orbitals (Fig. 1d). This may lead to a FM interlay coupling, akin to what was suggested for exchange couplings in $CrI_3$ bilayer [33,34]. Thus, to tune the AFM interlayer coupling to the FM one in MBT, partial $Mn^{2+}$ ions should be substituted by isovalent transition metal (TM) ions with empty $e_g$ orbitals.

For the choice of $TM^{2+}$ ions, it is vital to keep the intralayer FM ordering undamaged. This requires FM NN intralayer interactions between $Mn^{2+}$ and $TM^{2+}$ ions as well as between $TM^{2+}$ ions. Taking this into account, we select V as the substituent for the following reasons. First, similar to $Cr^{3+}$ ions in $CrI_3$ and $Cr_2Ge_2Te_6$ [35,36], $V^{2+}$ ions have FM NN intralayer interactions via the nealy-90° V-Te-V configuration, based on the Goodenough-Kanamori-Anderson rule [31,37,38]. Second, NN intralayer Mn and V also have a FM interaction for the reason discussed below. As shown in Fig. 1e-1f, the energy gains of FM ($E_{gain-FM}$) and AFM ($E_{gain-AFM}$) orderings are

$$E_{gain-FM} = t_1^2 / (\Delta + \Delta_V^c) \qquad (2),$$

and

$$E_{gain-AFM} = t_1^2 / (\Delta + \Delta_V^c + U_V) + t_2^2 / (\Delta_{Mn}^c + U_{Mn} - \Delta) \qquad (3).$$

In Eq. (2) and (3), $\Delta = E_{t_{2g}}^V - E_{t_{2g}}^{Mn}$, $\Delta_{Mn}^c = E_{e_g}^{Mn} - E_{t_{2g}}^{Mn}$ and $\Delta_V^c = E_{e_g}^V - E_{t_{2g}}^V$; $t_1$ ($t_2$) is the effective hopping between Mn-$t_{2g}$ and V-$e_g$ orbitals in the spin-up (spin-down) channels. Considering the strong correlation effect of 3$d$ electrons of Mn and V, i.e., large $U_{Mn}$ and $U_V$, the FM ordering have a larger energy gain than the AFM ordering.

Now, we demonstrate the validity of previous analyses through DFT calculations with three simple models. First, we construct a VBi$_2$Te$_4$ (VBT) SL to model the interactions between V ions and find that its NN intralayer interaction $J_1$ is indeed FM and much stronger than $J_2$ and $J_3$ (Table I). Second, a supercell of MBT SL with one Mn replaced by V is constructed to model the intralayer interactions between Mn and V, and FM NN Mn-V intralayer interaction is also confirmed (Table I). Lastly, we also obtain FM interlayer interactions between MBT and VBT SLs (Table I).

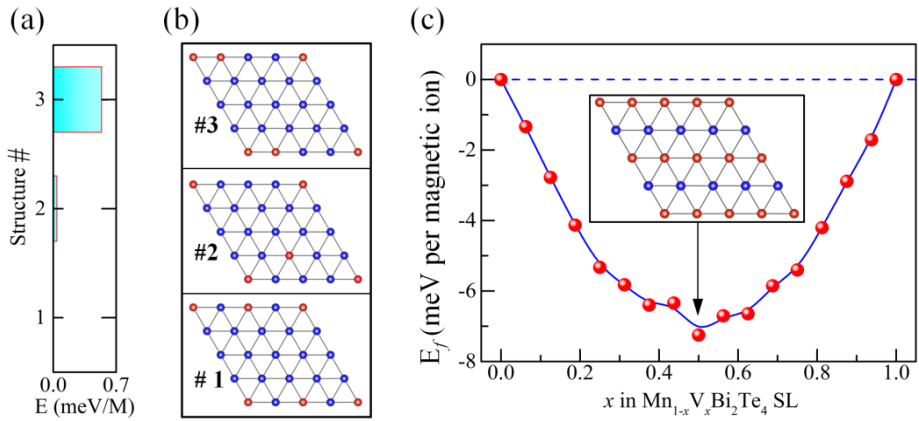

Fig. 2. (a) and (b) show the energies of different structures in Mn$_{1-x}$V$_x$Bi$_2$Te$_4$ SL with $x=0.125$. Structure #1 is set as reference. (c) Dependence of the formation energy of Mn$_{1-x}$V$_x$Bi$_2$Te$_4$ on V concentration $x$. The inset shows the lowest-energy distribution of Mn and V in Mn$_{0.5}$V$_{0.5}$Bi$_2$Te$_4$ SL.

To mimic the actual Mn$_{1-x}$V$_x$Bi$_2$Te$_4$ thin films, we consider a 4×4 supercell of MBT SL with $x$ from 0.0625 to 0.5. To reduce computational loads dealing with a large number of distribution configurations, structures are screened by means of the genetic algorithm (Table S5 in SM) [39]. As the case of $x=0.0625$ with only one V substituent per supercell is trivial, we begin with the case of $x=0.125$, namely, two V atoms in the supercell. As shown in Fig. 2a, the two V atoms tend to avoid each other as far as possible. This tendency is also clearly demonstrated especially in the cases of $x=0.1875$ and $x=0.25$ (Fig. S6 in SM). Overall, Mn and V distribute uniformly in Mn$_{1-x}$V$_x$Bi$_2$Te$_4$ SL, similarly to CrCl$_3$ and CrI$_3$ monolayers alloyed with tungsten [40,41]. Interestingly, Mn and V prefer to form an in-plane staggered stripe pattern in $Mn_{0.5}V_{0.5}Bi_2Te_4$ SL (inset in Fig. 2b).

To examine the feasibility of fabricating $Mn_{1-x}V_xBi_2Te_4$ in experiments, we calculate their formation energies ($E_f$) as

$$E_f\left(Mn_{1-x}V_xBi_2Te_4\right) = E_{Mn_{1-x}V_xBi_2Te_4} - (1-x)E_{MBT} - xE_{VBT} \qquad (4).$$

In Eq. (4), $E_{Mn_{1-x}V_xBi_2Te_4}$, $E_{MBT}$ and $E_{VBT}$ are energies of $Mn_{1-x}V_xBi_2Te_4$, MBT and VBT SLs, respectively. As shown in Fig. 2b, the formation energy as a function of V concentration $x$ indicates that $Mn_{1-x}V_xBi_2Te_4$ SL has a lower energy than the pure phases of MBT and VBT SLs. Especially, the case of $x=0.5$ has the lowest formation energy of -7.25 meV per magnetic atom. This indicates that $Mn_{1-x}V_xBi_2Te_4$ can be fabricated and stably exist without phase separation.

Let us now focus on the magnetic properties of $Mn_{0.5}V_{0.5}Bi_2Te_4$ with the in-plane staggered stripe pattern, since it has the lowest formation energy. Our calculations show that interlayer FM ordering has a lower energy than interlayer AFM ordering in the 2SL-$Mn_{0.5}V_{0.5}Bi_2Te_4$ model (Fig. 3a). Furthermore, the FM interlayer coupling persists in thicker $Mn_{0.5}V_{0.5}Bi_2Te_4$ films (Fig. 3a). Besides, $Mn_{0.5}V_{0.5}Bi_2Te_4$ thin films with thickness from 2 to 6 SLs all have positive magnetic anisotropy energies (MAEs) (Fig. 3b), i.e., an out-of-plane magnetic easy axis. So the AFM interlayer coupling of MBT is changed to the FM one in $Mn_{0.5}V_{0.5}Bi_2Te_4$ films by alloying with V.

The other vital issue for the use of $Mn_{0.5}V_{0.5}Bi_2Te_4$ as a quantum material is if its thickness-independent ferromagnetism can lead to robust QAHE. We first take 2-SL $Mn_{0.5}V_{0.5}Bi_2Te_4$ as an example. From Fig. 3d, we see that its band structure shows an insulator feature with a gap of 40.3 meV and, importantly, its Chern number is $C_N=-1$ when the Fermi level locates in the band gap (Fig. 3c). Accordingly, its ribbon shows a single 1D chiral edge state connecting the valance and conduction bands (Fig. 3i). These unequivocally evidence 2-SL $Mn_{0.5}V_{0.5}Bi_2Te_4$ as a Chern insulator, completely different from the zero plateau quantum anomalous Hall state in the pristine 2-SL AFM MBT [16]. More strikingly, $Mn_{0.5}V_{0.5}Bi_2Te_4$ thin films with thickness ranging from 3 to 6 SL are also Chern insulators with $C_N=-1$, with gaps of several tens of meV (Fig. 3e-3m). Therefore, QAHE can be easily realized in $Mn_{0.5}V_{0.5}Bi_2Te_4$ thin films without much sensitivity to the thickness.

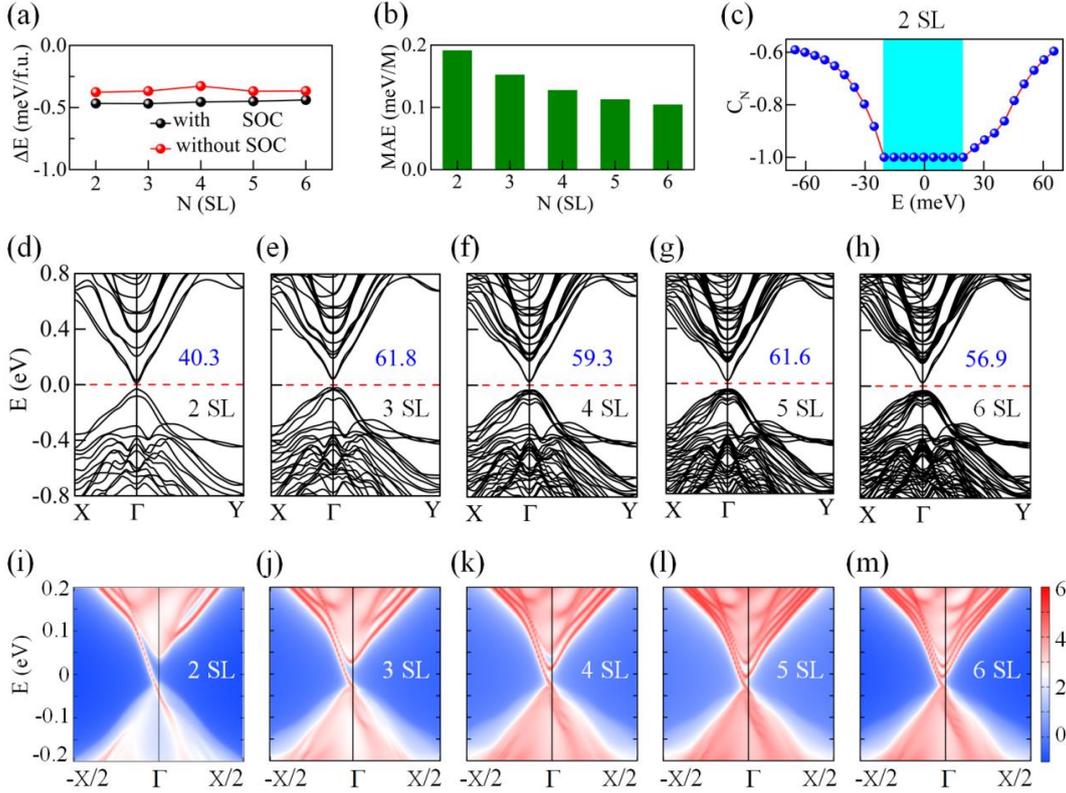

*Fig. 3. Magnetic and topological properties of $Mn_{0.5}V_{0.5}Bi_2Te_4$ thin films. (a) Energy difference $\Delta E=2(E_{FM}-E_{AFM})/(N-1)$ between interlayer FM ($E_{FM}$) and AFM ($E_{AFM}$) orderings as a function of thickness, N. (b) Dependence of MAE (meV per magnetic ion (M)) on thickness, N. (c) Dependence of $C_N$ on the position of Fermi level in 2-SL $Mn_{0.5}V_{0.5}Bi_2Te_4$. (d)-(h) Band structures of $Mn_{0.5}V_{0.5}Bi_2Te_4$ thin films with thickness from 2 to 6 SL. Blue numbers in (d)-(h) give band gaps in the unit of meV. (i)-(m) 1D chiral edge states of $Mn_{0.5}V_{0.5}Bi_2Te_4$ ribbons with thickness from 2 to 6 SL.*

We note that $Mn_{0.5}V_{0.5}Bi_2Te_4$ also has a much enhanced Curie temperature ($T_C$), compared with the pristine MBT. The NN intralayer exchange interactions of Mn-Mn,

Mn-V and V-V pairs in 2-SL $Mn_{0.5}V_{0.5}Bi_2Te_4$ are -1.49, -3.57 and -3.69 meV, respectively. Monte Carlo simulations based on the Heisenberg Hamiltonian, Eq. (1), as in previous studies of MBT thin films [4,16], show that the $T_C$ of $Mn_{0.5}V_{0.5}Bi_2Te_4$ thin films is above 42.1 K (see more details in Part VI in SM). Note that this $T_C$ is almost twice as high as $T_N$=23 K for the 5-SL MBT thin film [6]. Considering its band gap being also larger than 26 meV (corresponding to 300 K), it is perceivable that $Mn_{0.5}V_{0.5}Bi_2Te_4$ thin films can exhibit QAHE at a significantly higher temperature than the odd-SL MBT thin films.

In experiments, the as-grown MBT single crystals are electron doped [4,9,29,42,43], so it is often doped with Sb to compensate its n-type carriers [11,44]. It is therefore also useful to examine if the FM interlayer coupling and topological properties of $Mn_{0.5}V_{0.5}Bi_2Te_4$ are affected by Sb doping. To this end, we calculate $Mn_{0.5}V_{0.5}Bi_{2-2y}Sb_{2y}Te_4$ with $y$ in the vicinity of 0.3 at which $MnBi_{2-2y}Sb_{2y}Te_4$ has the n–p carrier transition [11]. As a result, $Mn_{0.5}V_{0.5}Bi_{2-2y}Sb_{2y}Te_4$ exhibits interlayer FM coupling and out-of-plane magnetic easy axis. However, it enters the Chern insulator phase from 3 SL rather than from 2 SL (Table S6 in SM).

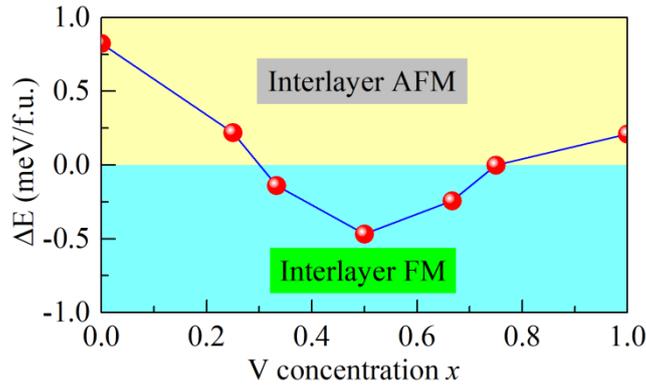

*Fig. 4. Phase diagram of the interlayer coupling in 2-SL $Mn_{1-x}V_xBi_2Te_4$ as a function of V concentration x. Energy difference $\Delta E=2(E_{FM}-E_{AFM})$ between interlayer FM ($E_{FM}$) and AFM ($E_{AFM}$) orderings is calculated with including spin orbit coupling.*

Finally, the FM interlayer coupling and thickness-independent QAHE exist with a wide range of V concentration $x$ in $Mn_{1-x}V_xBi_2Te_4$. As shown in Fig. 4, $Mn_{1-x}V_xBi_2Te_4$ has a FM interlayer coupling when V concentration $x$ is in the range from 0.333 to 0.667. Interestingly, the FM interlayer coupling is the strongest in $Mn_{0.5}V_{0.5}Bi_2Te_4$. When V concentration $x$ is in this range, our calculations indicate that $Mn_{1-x}V_xBi_2Te_4$ is in the Chern insulator phase without the dependence on its thickness (Table S7 in SM). The weak dependence of magnetic ordering on $x$ and the robust topological feature indicate that the requirement for controlling the growth condition for the synthesis of $Mn_{1-x}V_xBi_2Te_4$ samples is rather low. This makes $Mn_{1-x}V_xBi_2Te_4$ very attractive for producing QAHE materials in topological electronic applications.

In summary, based on systemic density functional theory calculations, we find a new strategy to tune the AFM interlayer coupling in $MnBi_2Te_4$ thin films to the FM one without damaging its nontrivial topology. By alloying about 50-50 vanadium and manganese, $Mn_{0.5}V_{0.5}Bi_2Te_4$ thin films have stable FM ground state, reasonably high Curie temperature, and manifest robust QAHE. This circumvents the need of applying an external magnetic field and monitoring the film thickness for quantum applications. Our findings offer a tempting route toward the development of emergent quantum materials for the development of quantum technologies, particularly for designs based on exotic topological features.

This work was supported by DOE-BES (Grant No. DEFG02-05ER46237) and DFT calculations were performed on parallel computers at NERSC supercomputer centers.

# References

[1] X.-L. Qi and S.-C. Zhang, Reviews of Modern Physics **83**, 1057 (2011).
[2] L. Fu and C. L. Kane, Phys. Rev. Lett. **100**, 096407 (2008).
[3] Y. Tokura, K. Yasuda, and A. Tsukazaki, Nature Reviews Physics **1**, 126 (2019).
[4] M. M. Otrokov, I. I. Klimovskikh, H. Bentmann, D. Estyunin, A. Zeugner, Z. S. Aliev, S. Gaß, A. U. B. Wolter, A. V. Koroleva, A. M. Shikin, M. Blanco-Rey, M. Hoffmann, I. P. Rusinov, A. Y.

Vyazovskaya, S. V. Eremeev, Y. M. Koroteev, V. M. Kuznetsov, F. Freyse, J. Sánchez-Barriga, I. R. Amiraslanov, M. B. Babanly, N. T. Mamedov, N. A. Abdullayev, V. N. Zverev, A. Alfonsov, V. Kataev, B. Büchner, E. F. Schwier, S. Kumar, A. Kimura, L. Petaccia, G. Di Santo, R. C. Vidal, S. Schatz, K. Kißner, M. Ünzelmann, C. H. Min, S. Moser, T. R. F. Peixoto, F. Reinert, A. Ernst, P. M. Echenique, A. Isaeva, and E. V. Chulkov, Nature **576**, 416 (2019).

[5] E. D. L. Rienks, S. Wimmer, J. Sánchez-Barriga, O. Caha, P. S. Mandal, J. Růžička, A. Ney, H. Steiner, V. V. Volobuev, H. Groiss, M. Albu, G. Kothleitner, J. Michalička, S. A. Khan, J. Minár, H. Ebert, G. Bauer, F. Freyse, A. Varykhalov, O. Rader, and G. Springholz, Nature **576**, 423 (2019).

[6] Y. Deng, Y. Yu, M. Z. Shi, Z. Guo, Z. Xu, J. Wang, X. H. Chen, and Y. Zhang, Science **367**, 895 (2020).

[7] D. Zhang, M. Shi, T. Zhu, D. Xing, H. Zhang, and J. Wang, Phys. Rev. Lett. **122**, 206401 (2019).

[8] J. Wu, F. Liu, M. Sasase, K. Ienaga, Y. Obata, R. Yukawa, K. Horiba, H. Kumigashira, S. Okuma, and T. Inoshita, Science advances **5**, eaax9989 (2019).

[9] S. H. Lee, Y. Zhu, Y. Wang, L. Miao, T. Pillsbury, H. Yi, S. Kempinger, J. Hu, C. A. Heikes, and P. Quarterman, Physical Review Research **1**, 012011 (2019).

[10] R. C. Vidal, H. Bentmann, T. R. F. Peixoto, A. Zeugner, S. Moser, C. H. Min, S. Schatz, K. Kißner, M. Ünzelmann, C. I. Fornari, H. B. Vasili, M. Valvidares, K. Sakamoto, D. Mondal, J. Fujii, I. Vobornik, S. Jung, C. Cacho, T. K. Kim, R. J. Koch, C. Jozwiak, A. Bostwick, J. D. Denlinger, E. Rotenberg, J. Buck, M. Hoesch, F. Diekmann, S. Rohlf, M. Kalläne, K. Rossnagel, M. M. Otrokov, E. V. Chulkov, M. Ruck, A. Isaeva, and F. Reinert, Physical Review B **100**, 121104 (2019).

[11] B. Chen, F. Fei, D. Zhang, B. Zhang, W. Liu, S. Zhang, P. Wang, B. Wei, Y. Zhang, Z. Zuo, J. Guo, Q. Liu, Z. Wang, X. Wu, J. Zong, X. Xie, W. Chen, Z. Sun, S. Wang, Y. Zhang, M. Zhang, X. Wang, F. Song, H. Zhang, D. Shen, and B. Wang, Nature Communications **10**, 4469 (2019).

[12] C.-Z. Chang, J. Zhang, X. Feng, J. Shen, Z. Zhang, M. Guo, K. Li, Y. Ou, P. Wei, and L.-L. Wang, Science **340**, 167 (2013).

[13] E. O. Lachman, A. F. Young, A. Richardella, J. Cuppens, H. Naren, Y. Anahory, A. Y. Meltzer, A. Kandala, S. Kempinger, and Y. Myasoedov, Science advances **1**, e1500740 (2015).

[14] I. Lee, C. K. Kim, J. Lee, S. J. Billinge, R. Zhong, J. A. Schneeloch, T. Liu, T. Valla, J. M. Tranquada, and G. Gu, Proceedings of the National Academy of Sciences **112**, 1316 (2015).

[15] C. Liu, Y. Wang, H. Li, Y. Wu, Y. Li, J. Li, K. He, Y. Xu, J. Zhang, and Y. Wang, Nature Materials (2020).

[16] M. M. Otrokov, I. P. Rusinov, M. Blanco-Rey, M. Hoffmann, A. Y. Vyazovskaya, S. V. Eremeev, A. Ernst, P. M. Echenique, A. Arnau, and E. V. Chulkov, Phys. Rev. Lett. **122**, 107202 (2019).

[17] J. Li, Y. Li, S. Du, Z. Wang, B.-L. Gu, S.-C. Zhang, K. He, W. Duan, and Y. Xu, Science Advances **5**, eaaw5685 (2019).

[18] P. E. Blochl, Physical Review B **50**, 17953 (1994).

[19] G. Kresse and D. Joubert, Physical Review B **59**, 1758 (1999).

[20] J. P. Perdew, A. Ruzsinszky, G. I. Csonka, O. A. Vydrov, G. E. Scuseria, L. A. Constantin, X. Zhou, and K. Burke, Phys. Rev. Lett. **100**, 136406 (2008).

[21] Z. S. Aliev, I. R. Amiraslanov, D. I. Nasonova, A. V. Shevelkov, N. A. Abdullayev, Z. A. Jahangirli, E. N. Orujlu, M. M. Otrokov, N. T. Mamedov, M. B. Babanly, and E. V. Chulkov, Journal of Alloys and Compounds **789**, 443 (2019).

[22] See Supplemental Material for Wyckoff positions of atoms and lattice constants of MBT bulk, effect of $U_{Mn}$ and $U_V$ on the magnetic and topological properties of $Mn_{1-x}V_xBi_2Te_4$, density of state and

Monte Carlo simulated Curie temperature $T_C$ of $Mn_{0.5}V_{0.5}Bi_2Te_4$, list of the configuration number of $Mn_{0.5}V_{0.5}Bi_2Te_4$ SL, Magnetic and topological properties of $Mn_{0.5}V_{0.5}Bi_{2-2y}Sb_{2y}Te_4$ with $y$ in the vicinity of 0.3 and $Mn_{1-x}V_xBi_2Te_4$ with $x$ from 0.333 to 0.667.

[23] Y. Hou and R. Wu, Nano letters **19**, 2472 (2019).
[24] Y. S. Hou, J. W. Kim, and R. Q. Wu, Physical Review B **101**, 121401 (2020).
[25] J. Klimeš, D. R. Bowler, and A. Michaelides, Journal of Physics: Condensed Matter **22**, 022201 (2009).
[26] J. Klimeš, D. R. Bowler, and A. Michaelides, Physical Review B **83**, 195131 (2011).
[27] A. A. Mostofi, J. R. Yates, Y.-S. Lee, I. Souza, D. Vanderbilt, and N. Marzari, Computer physics communications **178**, 685 (2008).
[28] Q. Wu, S. Zhang, H.-F. Song, M. Troyer, and A. A. Soluyanov, Computer Physics Communications **224**, 405 (2018).
[29] A. Zeugner, F. Nietschke, A. U. Wolter, S. Gaß, R. C. Vidal, T. R. Peixoto, D. Pohl, C. Damm, A. Lubk, and R. Hentrich, Chemistry of Materials **31**, 2795 (2019).
[30] Y. Li, Z. Jiang, J. Li, S. Xu, and W. Duan, Physical Review B **100**, 134438 (2019).
[31] J. Kanamori, Journal of Physics and Chemistry of Solids **10**, 87 (1959).
[32] H. Ehrenberg, M. Wiesmann, J. Garcia-Jaca, H. Weitzel, and H. Fuess, Journal of magnetism and magnetic materials **182**, 152 (1998).
[33] S. W. Jang, M. Y. Jeong, H. Yoon, S. Ryee, and M. J. Han, Physical Review Materials **3**, 031001 (2019).
[34] N. Sivadas, S. Okamoto, X. Xu, C. J. Fennie, and D. Xiao, Nano letters **18**, 7658 (2018).
[35] B. Huang, G. Clark, E. Navarro-Moratalla, D. R. Klein, R. Cheng, K. L. Seyler, D. Zhong, E. Schmidgall, M. A. McGuire, D. H. Cobden, W. Yao, D. Xiao, P. Jarillo-Herrero, and X. Xu, Nature **546**, 270 (2017).
[36] C. Gong, L. Li, Z. Li, H. Ji, A. Stern, Y. Xia, T. Cao, W. Bao, C. Wang, Y. Wang, Z. Q. Qiu, R. J. Cava, S. G. Louie, J. Xia, and X. Zhang, Nature **546**, 265 (2017).
[37] P. W. Anderson, Physical Review **79**, 350 (1950).
[38] J. B. Goodenough, Physical Review **100**, 564 (1955).
[39] D. C. Lonie and E. Zurek, Computer Physics Communications **183**, 690 (2012).
[40] F. Xue, Y. Hou, Z. Wang, and R. Wu, Physical Review B **100**, 224429 (2019).
[41] C. Huang, J. Feng, F. Wu, D. Ahmed, B. Huang, H. Xiang, K. Deng, and E. Kan, Journal of the American Chemical Society **140**, 11519 (2018).
[42] J. Cui, M. Shi, H. Wang, F. Yu, T. Wu, X. Luo, J. Ying, and X. Chen, Physical Review B **99**, 155125 (2019).
[43] J. Q. Yan, Q. Zhang, T. Heitmann, Z. Huang, K. Y. Chen, J. G. Cheng, W. Wu, D. Vaknin, B. C. Sales, and R. J. McQueeney, Physical Review Materials **3**, 064202 (2019).
[44] J.-Q. Yan, S. Okamoto, M. A. McGuire, A. F. May, R. J. McQueeney, and B. C. Sales, Physical Review B **100**, 104409 (2019).